\documentclass[12pt]{article}
\usepackage{graphicx}
\usepackage{setspace}
\usepackage{sidecap}
\def\pbnr{}
\def\speaker{Olga Bondarenko}
\def\onbehalfof{the BESIII collaboration}
\def\title{Recent Results on Isospin-Violating Transitions in Charmonium Studied with BESIII}
\def\affiliation{KVI, University of Groningen \\
Groningen, The Netherlands}
\def\support{The workshop was supported by the University of Manchester, IPPP, STFC, and IOP}
\newcommand\pubnumber{\pbnr}
\newcommand\pubdate{\today}
\def\Title#1{\begin{center} {\Large #1 } \end{center}}
\def\Author#1{\begin{center}{ \sc #1} \end{center}}

\newcommand{\OnBehalf}[1]{\sbox0{#1}\ifdim\wd0=0pt
        {}% if #1 is empty
	\else
	{\\on behalf of #1}% if #1 is not empty
	\fi}
\newcommand{\SupportedBy}[1]{\sbox0{#1}\ifdim\wd0=0pt
        {}% if #1 is empty
	\else
	{\footnote{#1}}% if #1 is not empty
	\fi}
\def\Address#1{\begin{center}{ \it #1} \end{center}}

\newcommand\pubblock{\includegraphics[width=5cm]{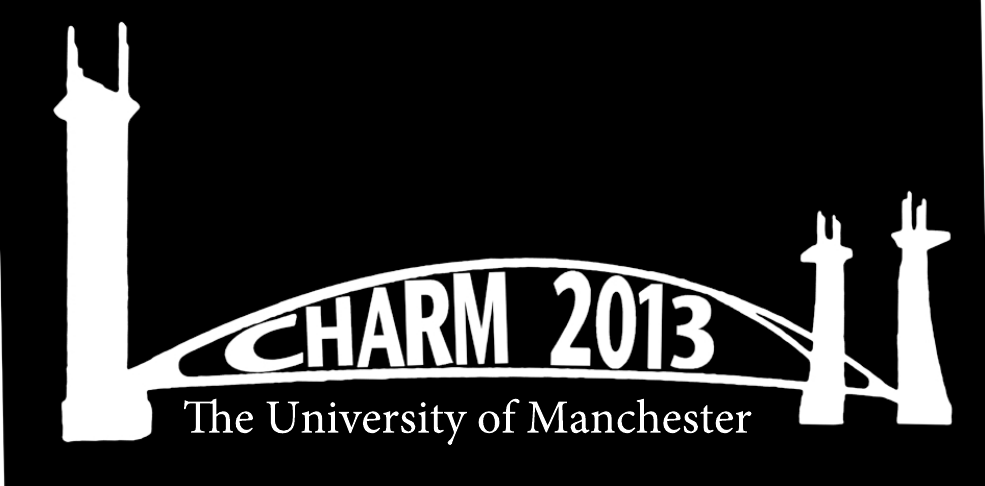}\hfill{\begin{tabular}{l} \pubnumber\\
         \pubdate  \end{tabular}}}
\newenvironment{Abstract}{\begin{quotation}  }{\end{quotation}}
\newenvironment{Presented}{\begin{quotation} \begin{center} 
             PRESENTED AT\end{center}\bigskip 
      \begin{center}\begin{large}}{\end{large}\end{center} \end{quotation}}

\def\venue{The 6$^{th}$ International Workshop on Charm Physics\\
(CHARM 2013)\\
Manchester, UK,  31 August -- 4 September, 2013}
\def\beq{\begin{equation}}
\def\eeq#1{\label{#1}\end{equation}}
\def\eeqn{\end{equation}}

\def\beqa{\begin{eqnarray}}
\def\eeqa#1{\label{#1}\end{eqnarray}}
\def\eeqan{\end{eqnarray}}

\let\bar=\overbar

\def\Dslash{\not{\hbox{\kern-4pt $D$}}}
\def\dslash{\not{\hbox{\kern-2pt $\del$}}}

\def\msb{{\bar{\ssstyle M \kern -1pt S}}}

\begin{document}
\begin{titlepage}
\pubblock

\vfill
\Title{\title}
\vfill
\Author{\speaker\SupportedBy{\support}\OnBehalf{\onbehalfof}}
\Address{\affiliation}
\vfill
\begin{Abstract}
Charmonium spectroscopy provides strong guidelines to effective field theories, thereby gaining insight into quark confinement and the generation of hadron masses. 
Hadronic transitions between charmonium states have been measured with unprecedented high precision with the BESIII spectrometer at IHEP Beijing, China. 
We present systematic studies of isospin-violating decays in charmonium, such as $\psi' \rightarrow \pi^{0} J/\psi$ and $\psi' \rightarrow \pi^{0} h_{c}$, 
and their improved measurements of their branching fractions. 
\end{Abstract}
\vfill
\begin{Presented}
\venue
\end{Presented}
\vfill
\end{titlepage}
\def\thefootnote{\fnsymbol{footnote}}
\setcounter{footnote}{0}

\section{Introduction}

The charmonium system covers a valuable energy region, where both the perturbative and non-perturbative QCD dynamics can be studied.  
Potential models have been successful in describing the charmonium spectrum, particularly below the open-charm threshold (for an overview, see~\cite{charmonium_overview}). 
However, in some cases experimental results differ significantly from the results of potential model calculations, 
and the possible explanation of these discrepancies are non-perturbative effects, such as effects of intermediate charmed-meson loops~\cite{quark_mass_hanhart}. 
The importance of these loops can be studied by the BESIII experiment. 

A potential model that is based on a non-relativistic framework and for which the potential is taken in the form of Coulomb plus linear terms, predicts a $2S$ hyperfine splitting of  
$M(\psi') - M(\eta_{c}') = 67.0$~MeV~\cite{eichten2004}. This is about 20~MeV larger than the experimental value of 48.5$\pm$3.3~MeV~\cite{hyperfine_bes3}. 
Including coupled-channels effects into the theoretical model improves the agreement between the theory and experiment, predicting a splitting with a size  
$M(\psi') - M(\eta_{c}') = 46.1$~MeV. It is tempting to conclude that the $\psi'-\eta_{c}'$ splitting reflects the influence of virtual decay channels~\cite{eichten2004}.   

Further hints to the importance of intermediate charmed-meson loops (ICML) can be in radiative transitions in charmonium. 
For instance, the prediction of a non-relativistic potential model for the width of the M1 radiative transition $\psi' \rightarrow \gamma \eta_{c}$ is 9.7~keV \cite{m1_theory}. 
This result is nearly one order of magnitude higher than the experimental value from the CLEO-c data, which gave a width of $\Gamma(\psi' \rightarrow \gamma \eta_{c}) = 0.97 \pm 0.14$~keV~\cite{m1_cleo}.  
Such a large discrepancy suggests that one cannot disregard ICML in theoretical calculations and that quenched quark models cannot be applied. 
Recently, a new unquenched theoretical model based on an effective Lagrangian approach was developed~\cite{m1_theory}. 
The prediction of this model is $\Gamma(\psi' \rightarrow \gamma \eta_{c}) = 2.05^{+2.65}_{-1.75}$~keV, which is in reasonable agreement with the measured value. 
Moreover, quenched lattice calculations predict a value $\Gamma(\psi' \rightarrow \gamma \eta_{c}) = 0.4\pm0.8$~keV \cite{m1_lattice} 
that overlaps within one standard deviation with the experimental result. 
Note that the large uncertainty in theoretical calculations complicates the physics interpretation and a further tuning of the existing models is needed.
From the experimental side, an improvement of the precision of the partial widths and, therefore, more measurements of radiative transitions in the charmonium mass region are needed.

The BESIII experiment~\cite{bes3} at the BEPCII $e^{+}e^{-}$ collider in Beijing obtained the world's largest data sets 
of vector charmonium states both below and above the open-charm threshold. 
The precision of various resonance parameters and decay properties of charmonium states has been improved significantly by BESIII 
with respect to previously published results. 
Furthermore, rare charmonium transitions have been successfully observed for the first time~\cite{hc_bes3}.

\begin{figure}[ht!]
 \begin{center}
 \includegraphics[scale=0.4]{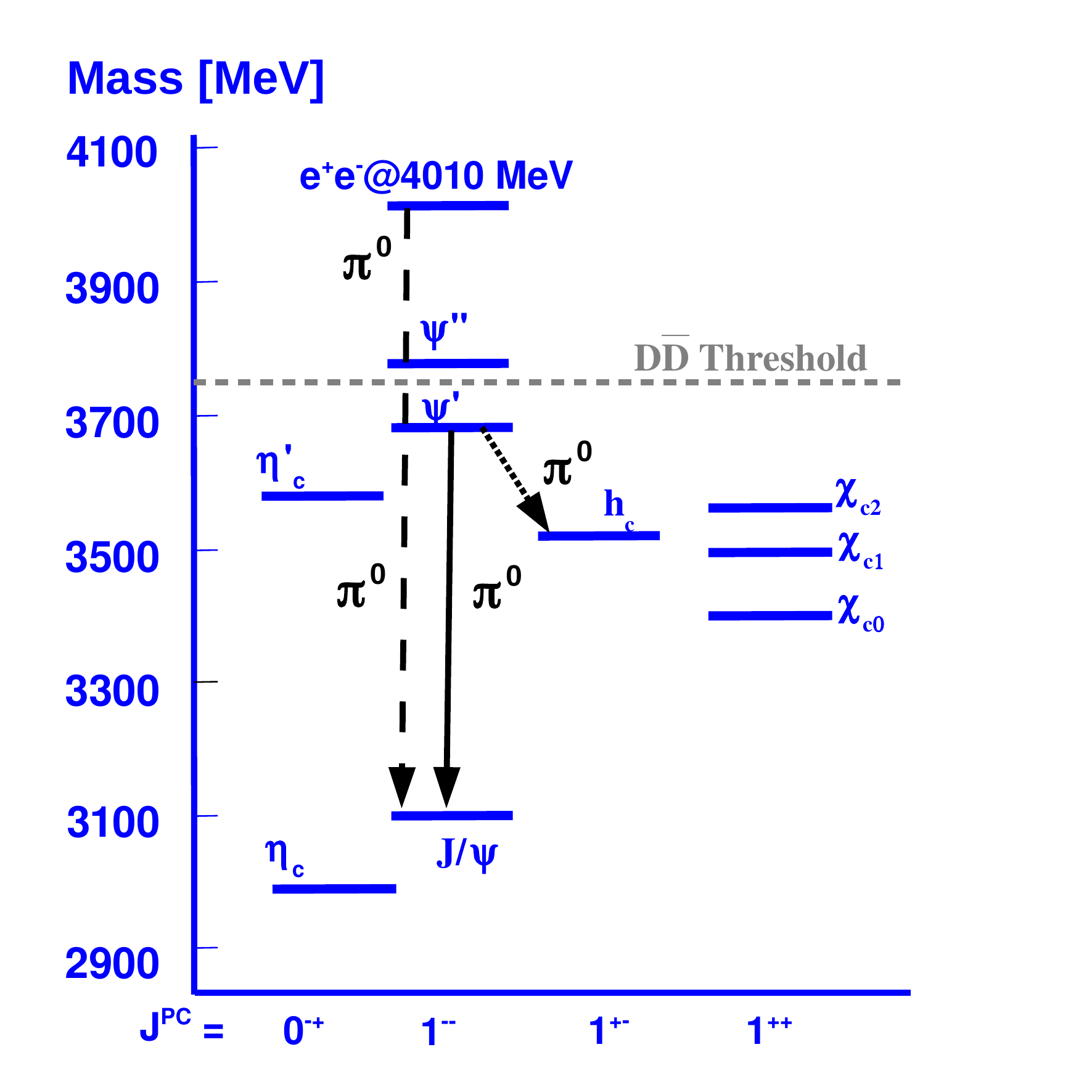}
     \caption{An overview of the studied isospin-violating transitions in the charmonium-mass region measured by BESIII. 
 	      The gray dashed arrow indicates the open-charm threshold. 
 	      The solid arrow indicates the transition $\psi'\rightarrow \pi^{0} J/\psi$. 
 	      The fine-dashed arrow shows the transition $\psi'\rightarrow \pi^{0} h_{c}$. 
 	      The dashed arrow shows the isospin-violating process $e^{+}e^{-} \rightarrow \pi^{0} J/\psi$ at a center-of-mass energy of $\sqrt s = 4010$~MeV.} 	    
     \label{fig:charm2013_levels}
 \end{center}
\end{figure}

In this paper, we present a fraction of the recently obtained results with BESIII, namely 
systematic studies of isospin-violating transitions in the charmonium system, such as $\psi' \rightarrow \pi^{0} J/\psi$ and 
$\psi' \rightarrow \pi^{0} h_{c}$. 
The relevant transitions in the charmonium system are depicted in Fig.~\ref{fig:charm2013_levels}.

% \begin{SCfigure}
%   \centering
%   \includegraphics[scale=0.38]{charm2013_levels.pdf}
%   \caption{An overview of the studied isospin-violating transitions in the charmonium-mass region measured by BESIII. 
%  	      The gray dashed arrow indicates the open-charm threshold. 
%  	      The solid arrow indicates the transition $\psi'\rightarrow \pi^{0} J/\psi$. 
%  	      The fine-dashed arrow shows the transition $\psi'\rightarrow \pi^{0} h_{c}$. 
%  	      The dashed arrow shows the isospin-violating process $e^{+}e^{-} \rightarrow \pi^{0} J/\psi$ at a center-of-mass energy of $\sqrt s = 4010$~MeV.
%  	      \label{fig:charm2013_levels}}
% \end{SCfigure}

%\section{Measurements of branching fractions of isospin-violating transitions in charmonium with BESIII}
%\label{sec:isospin}

%\section{}
\section{Measurements of branching fractions of \\ $\psi'\rightarrow \pi^{0}J/\psi$}

The hadronic transitions $\psi'\rightarrow \pi^{0}(\eta)J/\psi$ in the charmonium system were considered to be a reliable source for the extraction of the ratio between up and down quark masses~\cite{quark_mass}.
There are two possible sources of the isospin symmetry breaking, namely the electromagnetic processes and the difference between the masses of up and down quarks. 
Particularly for the isospin violating transition $\psi'\rightarrow \pi^{0}J/\psi$ (indicated by the solid arrow in Fig.~\ref{fig:charm2013_levels}), the  
contribution of electromagnetic processes was shown to be negligible ~\cite{em_effects}. Thus, for a long time, there was  hope to 
access the bare masses of light quarks using this transition. 
  
Based on a leading order QCD multipole expansion, the relation between the masses of the up and down quarks ($m_{u}, m_{d}$) and 
the ratio of the branching fractions of charmonium transitions $R_{\pi^{0}/\eta} = \frac{B(\psi'\rightarrow \pi^{0} J/\psi)}{B(\psi' \rightarrow \eta J/\psi)}$ can be written as~\cite{quark_mass}:

\begin{equation} 
    R_{\pi^{0}/\eta} = 3\left(\frac{m_{d} - m_{u}}{m_{d} + m_{u}}\right)^{2} \frac{F^{2}_{\pi^{0}}}{F^{2}_{\eta}} \frac{M^{4}_{\pi^{0}}}{M^{4}_{\eta}} \left | 
      \frac{\overrightarrow{q}_{\pi^{0}}}{\overrightarrow{q}_{\eta}}  \right | ^{3},
  \label{eq:eq1}
\end{equation}
where $F_{\pi^{0}(\eta)}$ and $M_{\pi^{0}(\eta)}$ are the decay constants and masses of the $\pi^{0}$ and $\eta$ mesons, respectively, 
and ${\overrightarrow{q}_{\pi^{0}(\eta)}}$ stands for the momentum of the corresponding particles produced in the transition $\psi' \rightarrow \pi^{0}(\eta)J/\psi$ in the $\psi'$ rest frame.

The up-down quark-mass ratio obtained using Eq.~\ref{eq:eq1} combined with the experimental results from the CLEO-c collaboration ~\cite{quark_mass_cleo} is 
$m_{u}/m_{d} = 0.40 \pm 0.01$, which is in discrepancy with other experimental methods, for example using the masses of the lightest scalar mesons ~\cite{quark_mass_light}.
It was suggested by various theoretical groups that the main source of discrepancy is the contribution of ICML~\cite{quark_mass_hanhart}. 
In Ref.~\cite{quark_mass_hanhart} it was shown that in the charmonium transitions $\psi' \rightarrow \pi^{0}(\eta) J/\psi$ the contributions of ICML are enhanced by a factor of two 
compared to the tree-level contributions. Thus, the light-quark mass ratio can only be extracted from these decays after establishing a complete effective field theory up to next-to-leading order. 

\begin{figure}[!htb]
  \begin{center}
    \includegraphics[scale=0.6]{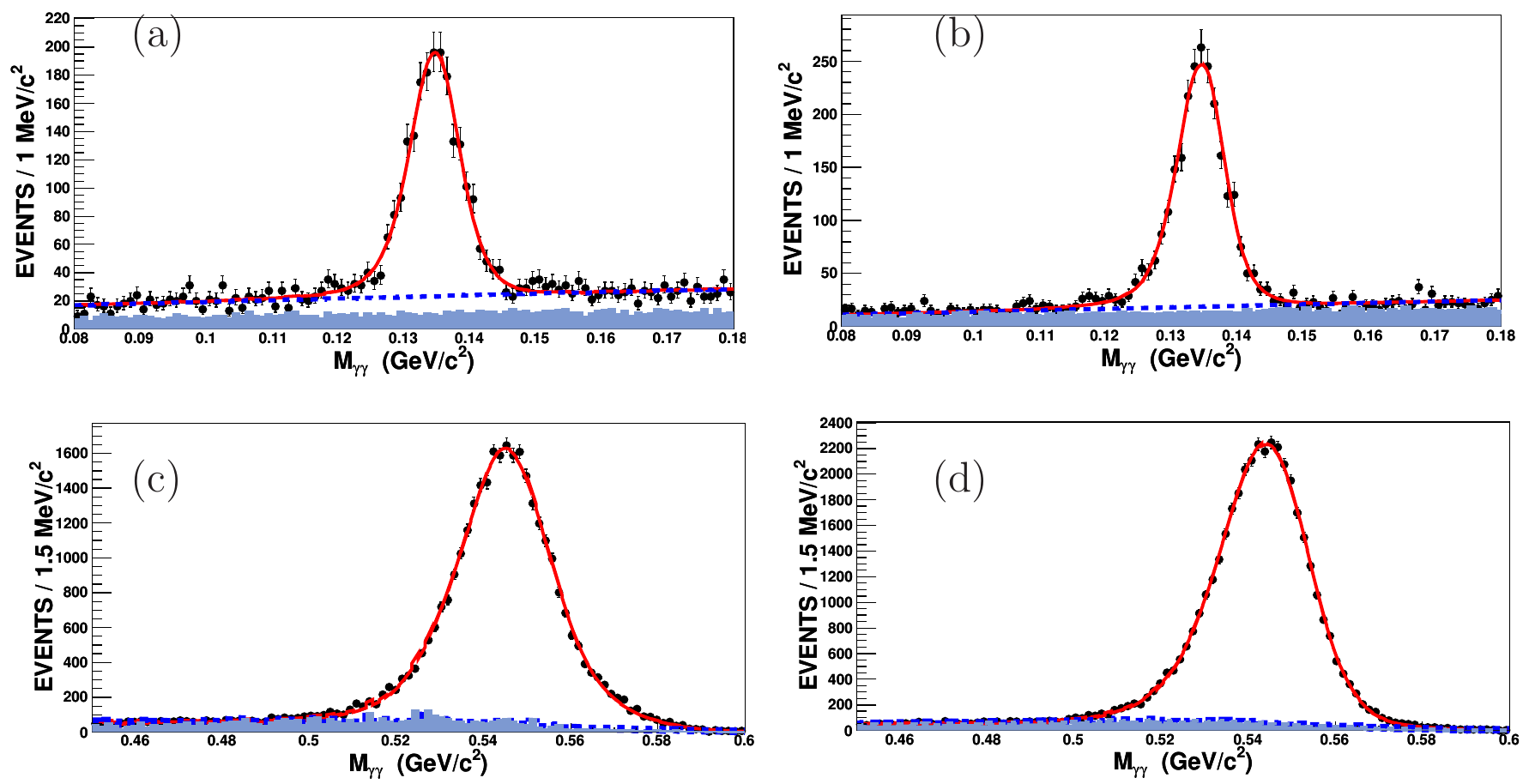}
    \caption{ 
	  Two-photon invariant-mass ($M_{\gamma\gamma}$) distributions and fit results. 
	  (a) $\psi'\to\pi^{0} J/\psi,J/\psi\to e^{+}e^{-}$, 
	  (b)$\psi'\to\pi^{0} J/\psi,J/\psi\to \mu^{+}\mu^{-}$,
	  (c)$\psi'\to\eta J/\psi,J/\psi\to e^{+}e^{-}$, 
	  (d)$\psi'\to\eta J/\psi,J/\psi\to \mu^{+}\mu^{-}$, 
	  where the points with error bars are data, and the red solid lines are the total fit results, 
	  and dashed lines are the fitted background shapes. 
	  The filled histograms present the shapes of the total background contributions obtained from a complete MC simulation 
	  of all possible $\psi'$ decays excluding the channel of interest, and from the side-bands analysis. \label{fig:mggfit}
	}
  \end{center}
\end{figure}

Recently, the BESIII collaboration performed a new measurement of the ratio of branching fractions 
$R_{\pi^{0}/\eta}$, with $\pi^{0}(\eta) \rightarrow \gamma\gamma$ and the leptonic decay mode of $J/\psi$ ($J/\psi \rightarrow e^{+}e^{-}/\mu^{+}\mu^{-}$) ~\cite{pi0jpsi_bes3}.
The reconstructed two-photon invariant-mass distributions of $\pi^{0}(\eta)$ together with the background contributions are shown in Fig.~\ref{fig:mggfit}. 
Clear signals from the $\pi^{0}$ and $\eta$ decays can be observed. Points with error bars represent data, the red solid lines represent total fit results, the dashed lines show the fitted background shapes and 
the filled histograms are the background shapes obtained from inclusive MC simulations of $\psi'$ decays and from a side-band analysis.
The value of the ratio $R_{\pi^{0}/\eta}$ was found to be 
$R_{\pi^{0}/\eta} = (3.74 \pm 0.06 \pm 0.04) \%$,
which is in good agreement with the previously reported measurements~\cite{quark_mass_cleo} and its precision is significantly improved.

The current estimate of the ratio of branching fractions 
$R_{\pi^{0}/\eta}$ provided by the non-relativistic field theories is $(11 \pm 6)\%$~\cite{quark_mass_hanhart}. 
This value falls within two standard deviations from the experimental result. 
The theoretical value lacks precision and the existing model needs further improvement. 
To this end, systematic experimental studies of isospin-violating transitions in the charmonium system are needed and can be carried out by the BESIII experiment. 
Such studies will in general help to improve our knowledge concerning non-perturbative effects in a relatively simple system such as charmonium, thereby testing the dynamics of QCD.

\section{Measurements of branching fractions of \\$\psi'\rightarrow \pi^{0}h_{c}$}

Another isospin-violating transition, $\psi' \rightarrow \pi^{0} h_{c}$ (shown with the fine-dashed line in Fig.~\ref{fig:charm2013_levels})  
was observed by the BESIII collaboration~\cite{hc_bes3}. 
This observation provided the first measurement of the branching fraction of this transition and its value was found to be 
$B(\psi'\rightarrow \pi^{0}h_{c}) = (8.4 \pm 1.3 \pm 1.0) \times 10^{-4}$, corresponding to a width of $\Gamma(\psi'\rightarrow \pi^{0} h_{c}) = 0.26 \pm 0.05$~keV.
Non-relativistic effective field theories predict that the contribution of the ICML is in the order of 10\% \cite{hc_hanhart}. 
By only accounting for the tree-level contributions and using a dimensional analysis, the width of the isospin-suppressed process is estimated to be 
$\Gamma(\psi' \rightarrow \pi^{0} h_{c}) = (0.9 \pm 0.6)C^{2}$~keV, where $C$ is the order of one. 
%The error is dominated by the error of the knowledge of the light quarks mass difference. (\cite{quark_mass_leutwyler}). 
The theoretical value is in reasonable agreement with the experimental data, which supports the model and the dominance of tree-level contributions.  
Also this value is consistent with predictions of 
single-channel calculations based on a QCD multipole expansion using the Cornell potential model, which gives a partial width of 
$\Gamma(\psi' \rightarrow \pi^{0} h_{c}) = (0.4 - 1.3)$~keV~\cite{hc_potential}.

\section{Measurements of branching fractions of \\$e^{+}e^{-} \rightarrow \pi^{0} J/\psi$ at $\sqrt{s}=$~4.009~GeV}

Effects of ICML above the open-charm threshold are expected to be significant according to 
existing theoretical models. 
For instance, an unquenched effective Lagrangian approach based on heavy-quark symmetry and chiral symmetry~\cite{pi0jpsi4010_theory} 
predicts that the isospin-violating process $e^{+}e^{-} \rightarrow \pi^{0} J/\psi$ is strongly affected by the open-charm effects and that 
the cross sections of these reactions as functions of energy may provide an opportunity for revealing the significance of ICML contributions.
Authors provide calculations of $\sigma(e^{+}e^{-}\rightarrow \pi^{0}J/\psi)$ for the energy range of 3.65-4.30~GeV.
Recently the isospin-breaking process $e^{+}e^{-} \rightarrow \pi^{0} J/\psi$ at $\sqrt{s} =$~4.009~GeV (
shown as the dashed line in Fig.~\ref{fig:charm2013_levels})  was studied by BESIII~\cite{pi0jpsi4010_bes3}. 
An upper limit on the $\pi^{0}J/\psi$ production cross section is set at $\sigma(e^{+}e^{-}\rightarrow \pi^{0}J/\psi) < 1.6$~pb at a 90\% confidence level. 
This measurement is in agreement with the upper limit set by the CLEO experiment~\cite{pi0jpsi4010_cleo}, 
$\sigma(e^{+}e^{-}\rightarrow \pi^{0}J/\psi) < 10$~pb at a 90\% confidence level and 
also does not contradict the theoretical prediction ~\cite{pi0jpsi4010_theory}. 
The theoretical calculations are dependent on the relative phases among the resonance transition amplitudes 
and give an upper limit of around $5\times 10^{-2}$~pb. 
However, more systematic measurements are needed to draw more solid conclusions.

\section{Summary}
In summary, we present recent BESIII results on a study of isospin-violating transitions in the charmonium-energy region, 
e.g. on the transitions $\psi' \rightarrow \pi^{0} J/\psi$, $\psi' \rightarrow \pi^{0} h_{c}$ and $e^{+}e^{-} \rightarrow \pi^{0} J/\psi$ at $\sqrt{s} =$~4.009~GeV. 
These experimental studies give an important input for the construction of an effective field theory that will reveal the effects of 
intermediate charmed-meson loops on transitions in charmonium system and would provide us more insight on the dynamics of non-perturbative QCD.

\singlespacing

\end{document}